\documentclass[preprint,aps,pre,amsmath,amssymb,amsfonts]{revtex4}
\usepackage{epsfig}

\begin{document}
\baselineskip18pt
\thispagestyle{empty}

\begin{center}\bf  Multiqubits Entanglement Based on Optimal Witness \end{center}
\vskip 1mm

\begin{center}
Ming-Jing Zhao$^{1}$, Zhi-Xi Wang$^{1}$ and Shao-Ming Fei$^{1,2}$

\vspace{2ex}

\begin{minipage}{5in}

\small $~^{1}$ {\small Department of Mathematics, Capital Normal
University, Beijing 100048}

{\small $~^{2}$ Institut f\"ur Angewandte Mathematik, Universit\"at Bonn, D-53115}

\end{minipage}

\vskip 4mm

(zhaomingjingde@126.com, wangzhx@mail.cnu.edu.cn,
     feishm@mail.cnu.edu.cn)

\end{center}

\vskip 2mm
\parbox{14cm}
{\footnotesize\quad We study the optimal entanglement witness with respect to
multiqubits $W$-states. We show such entanglement witnesses can be used to
distinguish genuine entangled states, different biseparable states
and fully separable states.}

\bigskip
\bigskip

\section{Introduction}
Quantum entanglement is of special significance in quantum
information processing and responsible for many quantum tasks such
as teleportation, dense coding, key distribution, error correction
etc. \cite{M.A.Nielsen}. Wether a state is entangled or not is one
of the most challenging open problems. For states of two-qubit or
one qubit and one qutrit systems, they are separable iff they are
PPT \cite{Peres A.,M. Horodecki}. For high dimensional and
multipartite systems, the separability problem becomes more
complicated. In particular experimentally realizable separability
criteria are less available. Besides the Bell's inequality
\cite{Sixia,Bao}, entanglement witness is the one that could be used
for experimental demonstration of quantum entanglement. For every
entangled state $\rho$ there exists an entanglement witness $W$ such
that $ {\rm Tr}(W \rho)<0$, and $ {\rm Tr}(W\sigma)\geq 0$ for all
separable states $\sigma$ \cite{M. Horodecki,B. M. Terhal,M.
Lewenstein,Philipp05,M. Lewenstein01}. A witness $W_1$ is said to be
finer than a witness $W_2$ iff $ {\rm Tr}(W_2 \rho)<0$ $\Rightarrow$
$ {\rm Tr}(W_1 \rho)<0$, i.e. $W_1$ detects all states that $W_2$
detects. A witness is optimal if there is no finer witness \cite{M.
Lewenstein}.

There are several methods to construct
entanglement witness \cite{S.Yu,O.,Cheng,Gza}. A universal
witness operator that detects the genuine entangled state $|\psi
\rangle$ is given by \cite{Mohamed,Philipp,Manfred}
\begin{equation}\label{max}
W=\alpha \mathbb{I} -|\psi \rangle \langle \psi |,
\end{equation}
where $\mathbb{I}$ is the identity operator and
\begin{equation}\label{maxb}
\alpha=\rm{max}_{|\phi \rangle \in B}|\langle \phi |\psi \rangle|^2,
\end{equation}
$B$ denotes the set of bipartite separable (biseparable) states.
Such witness can only detects genuine entangled states. Ref.\cite{Tzu}
proposed an optimal entanglement witness for certain
entangled state $|\psi \rangle$. They utilized the form (\ref{max}) with
the coefficient $\alpha$ given by
\begin{equation}\label{maxd}
\alpha=\rm{max}_{|\phi \rangle \in D}|\langle \phi |\psi \rangle|^2,
\end{equation}
where $D$ denotes the set of fully separable states. The witness of
the form (\ref{max}) with $\alpha$ given by (\ref{maxd}) is an
optimal entanglement witness for the pure state $|\psi \rangle$. In
this paper we construct a series of entanglement witnesses of form
(\ref{max}) that can be used to detect genuine entangled states, to
distinguish different biseparable states or other kinds of partially
separable states in multiqubits system.

An $N$-partite state $\rho$ is biseparable if it can be written as
\begin{equation}\label{def2sep}
 \rho=\sum_{perm\{i_1, i_2,
\cdots, i_N\}, m, j } p^{(j)}_{i_1, \cdots, i_N,m} \rho^{(j)}_{i_1,
i_2, \cdots, i_m} \otimes \rho^{(j)}_{i_{m+1}, \cdots, i_N},
\end{equation}
where $perm\{i_1, i_2, \cdots, i_N\}$ is a sum over all possible
permutations of the set of indices and $\sum_{perm\{i_1, i_2,
\cdots, i_N\}, m, j} p^{(j)}_{i_1, \cdots, i_N,m}=1$. Here
$\rho^{(j)}_{i_1, i_2, \cdots, i_m}$, $\rho^{(j)}_{i_{m+1}, \cdots,
i_N}$ are density
 matrices associated with the subsystems ${i_1, i_2, \cdots, i_m}$ and
${i_{m+1}, \cdots, i_N}$. We say that $\rho$ is genuine entangled if it
can not be written in the form of (\ref{def2sep}). If
$\rho$ is biseparable and can be further written as
\begin{equation}\label{def3sep}
\rho=\sum_{perm\{i_1, i_2, \cdots, i_N\}, m,s,j} p^{(j)}_{i_1,
\cdots, i_N,m,s} \rho^{(j)}_{i_1, i_2, \cdots, i_m} \otimes
\rho^{(j)}_{i_{m+1}, \cdots, i_{m+s}} \otimes \rho^{(j)}_{i_{m+s+1},
\cdots, i_N},
\end{equation}
$\rho$ is called tripartite separable. If $\rho$ is biseparable
but not tripartite separable, we say that it is a genuine
biseparable state. 4-partite separable states and genuine
tripartite separable states can be similarly defined in the sense that local actions
cannot increase the entanglement among the sub-quantum systems  \cite{V. Vedral}.

Therefore biseparable states are a convex subset $S_2$ of the whole
quantum states $S_1$. Tripartite separable states are a convex subset $S_3$
of biseparable states $S_2$, and so on. The ``smallest" convex
subset $S$ contains all the fully separable states that can be
written as
\begin{equation}\label{deftotalsep}
\rho=\sum_j p^{(j)} \rho_1^{(j)} \otimes \rho_2^{(j)} \otimes \cdots
\otimes \rho_N^{(j)}.
\end{equation}

For $N$-partite system there are $[\frac{N}{2}]$ kinds of
biseparable states, where $[\frac{N}{2}]$ is the largest integer
that smaller than or equal to $\frac{N}{2}$. We denote by $D_\alpha$
the set of biseparable states that $\alpha$ subsystems are separated
from the rest $N-\alpha$ subsystems,
\begin{eqnarray}
\rho=\sum_{perm\{i_1, i_2, \cdots, i_N\},j} p^{(j)}_{i_1, \cdots,
i_N} \rho^{(j)}_{i_1\cdots i_{\alpha}} \otimes
\rho^{(j)}_{i_{\alpha+1}, \cdots,
i_N}\label{def2sepn/2},~~~\alpha=1,2,...,[\frac{N}{2}].
\end{eqnarray}

\section{Witness for three-qubit system}
To analyze the entanglement properties we use the $W$-state to construct
entanglement witness. An $N$-qubit $W$-state $|W_N\rangle$ is given
by
\begin{equation}\label{nw}
|W_N \rangle = \frac{1}{\sqrt{N}} (|0 \cdots 01 \rangle + |0 \cdots
010 \rangle + \cdots +|10 \cdots 0\rangle).
\end{equation}

We first consider three-qubit case. As $\rm{max}_{|\phi \rangle \in
D}|\langle \phi |W_3 \rangle|^2 = \frac{4}{9}$ \cite{Tzu}, the
witness with respect to $|W_3\rangle$ has the form
\begin{equation}\label{w3w}
W_3=\frac{4}{9} \mathbb{I} -|W_3 \rangle \langle W_3 |.
\end{equation}
For any fully separable state $\rho$ one has
\begin{equation}\label{3fully}
 {\rm Tr}(W_3 \rho) \geq 0.
\end{equation}

$W_3$ has two different eigenvalues
$\frac{4}{9}$ and $-\frac{5}{9}$. Hence the
largest singular value of $W_3$ is $\frac{5}{9}$. That leads to
$|\langle \phi |W_3| \phi \rangle | \leq \frac{5}{9}$ for any pure
state $|\phi \rangle$, and
\begin{equation}\label{3any}
-\frac{5}{9}\leq  {\rm Tr}(W_3 \rho) \leq \frac{5}{9}
\end{equation}
for any mixed state $\rho$.

Now we view $|W_3 \rangle$ as a bipartite state. For example in the
bipartite decomposition of the first qubit and the last two qubits,
we have
\begin{eqnarray*}|W_3 \rangle &=& \sum_{i,j} C_{ij} |ij\rangle
=\frac{1}{\sqrt{3}}(|0\rangle \otimes |01\rangle + |0\rangle \otimes
|10\rangle+|1\rangle \otimes |00\rangle ),
\end{eqnarray*}
where coefficient matrix $C=(C_{ij})$ is given by
\begin{eqnarray*} C= \frac{1}{\sqrt{3}}\left(
\begin{array}{cccc}
0 & 1 & 1 & 0 \\
1 & 0 & 0 & 0
\end{array}
\right).\\[3mm]
\end{eqnarray*}
The largest singular value of $C$ is $\sqrt{\frac{2}{3}}$, which is
easily verified to be independent of the ways of detailed bipartite
decompositions. Thus for any pure state $|\phi \rangle$ in $D_1$ of
three-qubit system, one has $|\langle \phi |W_3 \rangle| \leq
\sqrt{\frac{2}{3}}$ \cite{Mohamed}. Therefore for arbitrary state
$\rho$ in $D_1$ of three-qubit system, we have
\begin{equation}\label{32sep}
-\frac{2}{9} \leq {\rm Tr}(W_3 \rho) \leq \frac{5}{9}.
\end{equation}

From above we can conclude that if $\rho$ is fully separable, it
satisfies (\ref{3fully}). If $\rho$ is biseparable, it satisfies
(\ref{32sep}). And if ${\rm Tr}(W_3 \rho)\leq -\frac{2}{9}$, $\rho$
is genuine entangled.

{\sf Remark}~If one chooses $\alpha$ in (\ref{maxb}) to construct
the witness (\ref{max}), then in stead of $W_3$, one obtains a witness
$\overline{W}_3=\frac{2}{3}-|W_3\rangle \langle
W_3|$\cite{Mohamed,Tzu}. It is obvious that $W_3$ detects
entanglement better than $\overline{W}_3$ does. In \cite{Mohamed} it
is found that $\overline{W}_3$ can detect the genuine entanglement
of the mixture of $|W_3\rangle$ and white noise $\rho=p|W_3\rangle
\langle W_3|+\frac{(1-p)}{8}\mathbb{I}$ up to $p>\frac{13}{21}$. By
using $W_3$ one can see that $\rho$ is genuine entangled for
$p>\frac{13}{21}$, and entangled for $p>\frac{23}{63}$. In fact
$W_3$ is tangent to $S$ with the tangent point between $W_3$ and
$S$: $(\sqrt{\frac{2}{3}}|0\rangle
+\sqrt{\frac{1}{3}}|1\rangle)^{\otimes 3}$. While $\overline{W} _3$
is tangent to $S_2$ with the tangent plane between $\overline{W} _3$
and $S_2$: $\rho=\sum_{i=1}^3 p_i |\psi _i\rangle \langle \psi_i |$,
where $\sum_{i=1}^3p_i=1$,
$|\psi_1\rangle=\frac{1}{\sqrt{2}}(|001\rangle+|010\rangle)$,
$|\psi_2\rangle=\frac{1}{\sqrt{2}}(|001\rangle+|100\rangle)$,
$|\psi_3\rangle=\frac{1}{\sqrt{2}}(|010\rangle+|100\rangle)$, see Fig. 1.

\begin{figure}
\begin{center}
\resizebox{7cm}{!}{\includegraphics[angle=90]{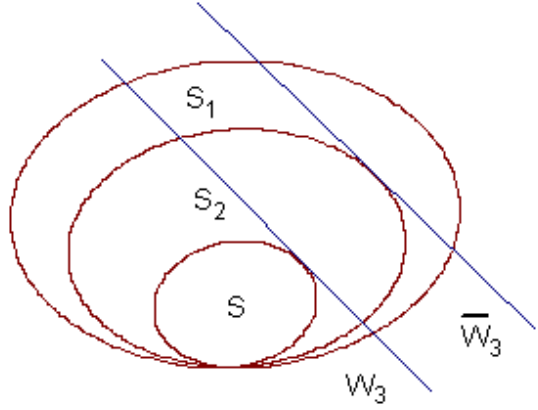}}\caption{$W_3$
is tangent to $S$ with a tangent point and $\overline{W}_3$ is
tangent to $S_2$ with a tangent plane.}
\end{center}
\end{figure}

For tripartite case, a measure of genuine entanglement of tripartite
states, called three-tangle $\tau_3(\rho)$, has been
introduced Ref. \cite{V. Coffman}.
If a three-qubit state $\rho$ is biseparable or fully
separable, then $\tau_3(\rho)=0$. For $\rho=p|W_3\rangle\langle W_3|
+ (1-p)|\rm GHZ\rangle\langle GHZ|$, it is showed \cite{R. Lohmayer}
that when $0\leq p\leq 0.373$, $\rho$ is genuine entangled. But $W_3$ can
not detect any entanglement for such genuine entangled states. While
for $0.667 \leq p \leq 1$, $\rho$ is genuine entangled which can be
detected by $W_3$ but not by the three-tangle. Thus we can conclude that
the witness and the three-tangle may detect different genuine entangled states.

Let us now analyze which kinds of pure entangled states that can be
detected by $W_3$. For a general three-qubit state  $|\phi \rangle =
\sum_{i,j,k=0}^1 a_{ijk}|ijk\rangle$, with $\sum_{i,j,k=0}^1
|a_{ijk}|^2=1$, one has ${\rm Tr}(W_3|\phi \rangle \langle
\phi|)=\frac{4}{9}-\frac{1}{3}|a_{001}+a_{010}+a_{100}|^2$.
Therefore $|\phi\rangle$ is entangled if
$|a_{001}+a_{010}+a_{100}|>\frac{2}{\sqrt{3}}$ and genuine entangled
if $|a_{001}+a_{010}+a_{100}|>\sqrt{2}$. In fact any three-qubit
pure state can be written as $|\psi \rangle = \lambda_0 |000\rangle
+\lambda_1
e^{i\theta}|100\rangle+\lambda_2|101\rangle+\lambda_3|110\rangle+\lambda_4|111\rangle$,
where $\lambda_i\geq0$, $\sum_i\lambda_i^2=1$, $\theta \in
[0,\pi]$\cite{A.Ac}. $|\psi \rangle$ can be detected by $W_3$ as
entangled if $\lambda_1>\frac{2}{3}$ and as genuine entangled if
$\lambda_1>\frac{\sqrt{6}}{3}$.

\section{Witness for N-qubit system}
Now we study the witness
with respect to $|W_N \rangle$ for $N$-qubit system. As $max_{|\phi
\rangle \in D}|\langle \phi |W_N \rangle|^2 =
(\frac{N-1}{N})^{N-1}$, we get the related witness
\begin{equation}\label{new}
W_N = (\frac{N-1}{N})^{N-1}\mathbb{I}-|W_N \rangle \langle W_N|.
\end{equation}

$W_N$ has two different eigenvalues $(\frac{N-1}{N})^{N-1}$ and
$(\frac{N-1}{N})^{N-1}-1$. Since $\{(\frac{N-1}{N})^{N-1}\}$ is a
decreasing sequance and for $N=2$,
$(\frac{N-1}{N})^{N-1}=\frac{1}{2}$, we have $(\frac{N-1}{N})^{N-1} \leq
\frac{1}{2}$ for $N\geq 2$. Hence $|(\frac{N-1}{N})^{N-1}| \leq |
(\frac{N-1}{N})^{N-1}-1| $ and the largest singular value of $W_N$
is $1-(\frac{N-1}{N})^{N-1}$. Therefore for any pure state $|\phi
\rangle$, $| \langle \phi |W_N| \phi \rangle | \leq 1-
(\frac{N-1}{N})^{N-1}$ and for arbitrary mixed state $\rho$,
\begin{equation}\label{nany}
(\frac{N-1}{N})^{N-1}-1 \leq {\rm Tr}(W_N \rho) \leq
1-(\frac{N-1}{N})^{N-1}.
\end{equation}
Or explicitly for arbitrary state $\rho$ and $N\geq2$, $-0.6322\leq
{\rm Tr}(W_N \rho)\leq0.6322$ as
$\lim_{N\rightarrow+\infty}(\frac{N-1}{N})^{N-1}=\frac{1}{e}\doteq
0.36788$.

We now focus on bipartite decompositions of $N$-qubit systems. If we
consider $|W_N \rangle$ as a bipartite state in $1|2\cdots N$
system, $|W_N \rangle=\sum_{i,j} C^{(1)}_{ij} |ij\rangle$, where
$C^{(1)}=(C^{(1)}_{ij})$ satisfies
\begin{eqnarray*} C^{(1)}C^{(1)\dag}= \frac{1}{N}\left(
\begin{array}{cc}
N-1 & ~~0  \\
0 & ~~1
\end{array}
\right).
\end{eqnarray*}
The largest singular value of the coefficient matrix $C^{(1)}$ is
$\sqrt{\frac{N-1}{N}}$, which is again the same for all possible bipartite
decompositions in $D_1$. Therefore for any pure state $|\psi
\rangle$ in $D_1$ of $N$-qubit system, one has $|\langle\psi
|W_N\rangle| \leq \sqrt{\frac{N-1}{N}}$. For mixed state $\rho$ in
$D_1$,
\begin{equation}\label{n1bi}
(\frac{N-1}{N})^{N-1}-\frac{N-1}{N}\leq {\rm Tr}(W_N \rho)\leq 1-
(\frac{N-1}{N})^{N-1}.
\end{equation}

Generally if we view $|W_N \rangle$ as a bipartite state in
$12\cdots k|k+1\cdots N$ system ($1\leq k\leq[\frac{N}{2}]$), we
have
$|W_N \rangle=\sum_{i,j} C^{(k)}_{ij} |ij\rangle$
and
\begin{eqnarray*} C^{(k)}C^{(k) \dag}= \frac{1}{N}\left(
\begin{array}{cccccccccc}
N-k & 0 & 0 & 0 & \cdots & 0 & 0 & \cdots & 0 & 0\\
0 & 1 & 1 & 0 & \cdots & 1 & 0 & \cdots & 1 & 0\\
0 & 1 & 1 & 0 & \cdots & 1 & 0 & \cdots & 1 & 0\\
0 & 0 & 0 & 0 & \cdots & 0 & 0 & \cdots & 0 & 0\\
\cdots & \cdots & \cdots & \cdots & \cdots & \cdots & \cdots & \cdots & \cdots & \cdots
\\
 0 & 1 & 1 & 0 & \cdots & 1 & 0 & \cdots & 1 & 0\\
 0 & 0 & 0 & 0 & \cdots & 0 & 0 & \cdots & 0 & 0\\
 \cdots & \cdots & \cdots & \cdots & \cdots & \cdots & \cdots & \cdots & \cdots & \cdots
 \\
 0 & 1 & 1 & 0 & \cdots & 1 & 0 & \cdots & 1 & 0\\
 0 & 0 & 0 & 0 & \cdots & 0 & 0 & \cdots & 0 & 0
\end{array}
\right)_{2^k \times 2^k},
\end{eqnarray*}
where $C^{(k)}=(C^{(k)}_{ij})$. Except for the first column (resp.
row), there are $k$ entries of $1$ in each column (resp. row). The
largest singular value of the coefficient matrix $C^{(k)}$ is
$\sqrt{\frac{N-k}{N}}$. Similarly we have for any pure state $|\psi
\rangle$ in $D_k$ of an $N$-qubit system, $|\langle\psi |W_N\rangle|
\leq \sqrt{\frac{N-k}{N}}$. As
$1-(\frac{N-1}{N})^{N-1} < (\frac{N-1}{N})^{N-1} + \frac{N-k}{N}$
for $ 1\leq k \leq [\frac{N}{2}]$, for any mixed state $\rho$
in $D_k$, we obtain
\begin{equation}
\label{nkbi} (\frac{N-1}{N})^{N-1}-\frac{N-k}{N}\leq {{\rm Tr}(W_N
\rho)}\leq 1- (\frac{N-1}{N})^{N-1}.
\end{equation}

In summary we have that if an $N$-qubit state $\rho$ is fully
separable, then ${\rm Tr}(W_N\rho)\geq 0$. If $\rho$ is biseparable,
then (\ref{n1bi}) holds, which is true for any biseparable states as
$\frac{N-k}{N}\leq 1$. Moreover if $\rho$ is not biseparable in
$D_k~(1\leq k<[\frac{N}{2}])$, then $\rho$ is not biseparable in
$D_{k+1}$. In particular, if $\rho$ is not biseparable in $D_1$,
then $\rho$ is genuine entangled, namely if ${\rm
Tr}(W_N\rho)<(\frac{N-1}{N})^{N-1}-\frac{N-1}{N}$, then $\rho$ is
genuine entangled.

As an example we consider the state $\rho = p |W_N \rangle \langle
W_N | + (1-p)|\rm GHZ_N \rangle \langle \rm GHZ_N |$, $0\leq p\leq
1$, where $|\rm GHZ_N\rangle=\frac{1}{\sqrt{2}}(|00\cdots 0 \rangle
+|11\cdots 1\rangle)$. It is direct to verify that ${\rm
Tr}(W_N\rho)=(\frac{N-1}{N})^{N-1}-p$. That is, $\rho$ is entangled
for $(\frac{N-1}{N})^{N-1}<p\leq1$ and genuine entangled for
$\frac{N-1}{N}<p\leq1$. For state $\rho=p|W_N \rangle \langle W_N
|+\frac{1-p}{2^N}\mathbb{I}$, one has that $\rho$ is entangled for
$(2^N(\frac{N-1}{N})^{N-1}-1)/(2^N-1)<p\leq1$ and genuine entangled
for $(2^N(\frac{N-1}{N})-1)/(2^N-1)<p\leq1$. As $\lim_{N \rightarrow
\infty} (2^N(\frac{N-1}{N})^{N-1}-1)/(2^N-1)=\frac{1}{e}$, $\lim_{N
\rightarrow \infty}(2^N(\frac{N-1}{N})-1)/(2^N-1)=1$, one gets that
the larger $N$ is, the less genuine entanglement $W_N$ can detect.

For a general pure $N$-qubit state, $|\phi \rangle = \sum
_{i_1i_2\cdots i_N=0}^1 a_{i_1i_2\cdots i_N}|i_1i_2\cdots
i_N\rangle$ with $\sum _{i_1i_2\cdots i_N=0}^1 |a_{i_1i_2\cdots
i_N}|^2=1$, one has ${\rm Tr}(W_N|\phi \rangle \langle
\phi|)=(\frac{N-1}{N})^{N-1}-\frac{1}{N}|a_{0\cdots01}+a_{0\cdots10}+\cdots+a_{10\cdots0}|^2$.
Therefore $|\phi \rangle$ is entangled if
$|a_{0\cdots01}+a_{0\cdots10}+\cdots+a_{10\cdots0}|>(N-1)^{\frac{N-1}{2}}
N^{\frac{2-N}{2}}$ and genuine entangled if
$|a_{0\cdots01}+a_{0\cdots10}+\cdots+a_{10\cdots0}|>\sqrt{N-1}$.
Thus the entanglement of the states in these parameter region can be
detected experimentally by the the witness $W_N$.

\section{Conclusions}
We have investigated the entanglement
properties of multiqubits states by using the optimal entanglement
witness with respect to $|W_N\rangle$, $N\geq3$. These witnesses can
be used to distinguish genuine entangled states, different
biseparable states in $S_2$ and fully separable states
experimentally. The entanglement properties for states in $S_3$,
$S_4$,... can be also similarly studied based on optimal entanglement
witness.

\bigskip

\noindent{\bf Acknowledgments}
 Project supported by the NSF of
Beijing(No. 1092008) and Beijing Municipal Commission of Education
(KZ200810028013, KM200810028003).

\end{document}